\documentclass[useAMS,usenatbib,usegraphicx]{mn2e}
\usepackage{amsmath}
\usepackage[normalem]{ulem}
\usepackage{graphicx,color}
\usepackage{times,subfigure,rotating,tikz}
\def\msun{$M_{\odot}$}
\def\hi{\mbox{H\sc{i}}}

\newcommand{\gtsimeq}{\raisebox{-0.6ex}{$\,\stackrel{\raisebox{-.2ex}{$\textstyle >$}}{\sim}\,$}}
\title[Gas accretion/depletion timescales in ETGs]
{On the depletion and accretion timescales of cold gas in local early-type galaxies}
\author[T.\ A.\ Davis \& M.\ Bureau] {Timothy A.\
  Davis$^{1,2}$\thanks{E-mail: t.davis4@herts.ac.uk}
  and Martin Bureau$^{3}$\\
  $^{1}$Centre for Astrophysics Research, University of Hertfordshire,
  Hatfield, Herts AL1 9AB, UK\\
  $^{2}$School of Physics \&\ Astronomy, Cardiff University, Queens Buildings, The Parade, Cardiff, CF24 3AA, UK\\
  $^3$Sub-department of Astrophysics, University of Oxford, Denys
  Wilkinson Building, Keble Road, Oxford OX1~3RH, UK}

\begin{document}
\maketitle
%
%
\begin{abstract}
  We consider what can be learnt about the processes of gas accretion
  and depletion from the kinematic misalignment between the cold/warm
  gas and stars in local early-type galaxies.  Using simple analytic
  arguments and a toy model of the processes involved, we show that
  the lack of objects with counter-rotating gas reservoirs strongly
  constrains the relaxation, depletion and accretion timescales of gas
  in early-type galaxies. Standard values of the accretion rate, star
  formation efficiency and relaxation rate are not simultaneously
  consistent with the observed distribution of kinematic
  misalignments. To reproduce that distribution, both fast gas
  depletion ($t_{\rm dep}\,\la\,10^8$~yr; e.g.\ more efficient star
  formation) and fast gas destruction (e.g.\ by active galactic
  nucleus feedback) can be invoked, but both also require a high rate
  of gas-rich mergers ($>1$~Gyr$^{-1}$). Alternatively, the relaxation
  of misaligned material could happen over very long timescales
  ($\simeq100$ dynamical times or $\approx1$-$5$~Gyr). We explore the
  various physical processes that could lead to fast gas depletion
  and/or slow gas relaxation, and discuss the prospects of using
  kinematic misalignments to probe gas-rich accretion processes in the
  era of large integral-field spectroscopic surveys.
\end{abstract}
%
%
\nobreak
\section{Introduction}
\label{sec:intro}

In all its forms, gas is one of the most important drivers of galaxy
evolution. The availability of cold gas (atomic and molecular)
regulates star formation, and thus the evolution and fate of galaxies,
from the beginning of the universe to today. Many factors affect the
supply and removal of gas. Some processes allow gas to flow onto
galaxies and form stars (e.g.\ cold accretion, mergers, hot halo
cooling and stellar mass loss; e.g.\
\citealt{2005MNRAS.363....2K,Parriott:2008co,Bauermeister:2010jl}),
while others compete to remove, destroy, and/or prevent gas accretion
(e.g.\ stellar and active galactic nucleus (AGN) feedback, outflows,
the development of hot halos and virial shocks; e.g.\
\citealt{1998A&A...331L...1S,Birnboim:2003fo,2006MNRAS.365...11C}). The
balance reached by these processes determines if a galaxy forms stars
throughout its life, or if it becomes a gas-starved ``red and dead''
object. Understanding these processes is thus vital to further develop
theories of galaxy evolution.

One laboratory where we are able to study these processes at work is
gas-rich early-type galaxies (ETGs; ellipticals and lenticulars). ETGs
typically have red optical colours, forming a ``red sequence'' in
optical colour-magnitude diagrams
\citep[e.g.][]{2004ApJ...600..681B}. {The majority of the stars in current-day ETGs} were
already in place at $z=2$ \citep[e.g.][]{Bower:1992un,2005ApJ...621..673T}, suggesting
that some of the physical processes listed above have kept them
relatively gas-free ever since.
 
Nevertheless, a small proportion of ETGs do have cold gas and
associated star formation, consistent with a recent regeneration of
their gas reservoir \citep{2014MNRAS.444.3408Y}. {Such low level activity has been found to contribute up-to 30\% of the mass of some ETGs since $z=1$ \citep[e.g.][]{2014MNRAS.437L..41K}}. 

A great deal of work
has in fact been done in recent years to quantify the amount of gas
present in the ETG population. Studies of statistically complete
samples have shown that $\approx25\%$ of ETGs have $>10^7$~\msun\ of
cold molecular gas
\citep[e.g.][]{2007MNRAS.377.1795C,Welch:2010in,2011MNRAS.414..940Y},
while $\approx40$\% have sizeable atomic gas reservoirs {($>10^7$~\msun)}
\citep[e.g.][]{2006MNRAS.371..157M,Sage:2006ko,2012MNRAS.422.1835S}. Dust
is also present in a large fraction of ETGs
\citep[e.g.][]{2012ApJ...748..123S}, and the presence of gas has been
inferred from low-level residual star formation
\citep[e.g.][]{1989ApJS...70..329K,2005ApJ...619L.111Y,2007ApJS..173..619K,2010ApJ...714L.290S,2009ApJ...695....1T,2010MNRAS.402.2140S}.

A quenched object has only two paths by which to rebuild its cold ISM:
an internal one (cooling of material lost from stars during their
evolution) and an external one (wet mergers, cold accretion, and/or
cooling of hot/shocked accreted gas). Stellar mass loss must of course
always be present, but the cold gas detection rate smaller than unity
suggests that the majority of stellar ejecta does not cool but rather
remains warm or hot (e.g.\ settling at the virial temperature of the
galaxy; e.g. \citealt{2008ApJ...681.1215P}). Various observational studies have suggested that the external channel dominates in field environments.
The evidence for this comes from the large fraction of gas-rich ETGs
that are morphologically disturbed in deep imaging
\citep[e.g.][]{2005AJ....130.2647V,2014arXiv1410.0981D}, the presence
of young kinematically decoupled cores in IFU surveys
\citep[e.g.][]{McDermid:2006bj}, the orders of magnitude surplus of
interstellar medium (ISM) compared to expectations from stellar mass
loss
\citep[e.g.][]{1998A&A...338..807M,1989ApJS...70..329K,2012MNRAS.419.2545R,2012MNRAS.423...49K},
the large spread of gas-to-dust ratios within their cold ISM
\citep{2012ApJ...748..123S,2015MNRAS.449.3503D} and the kinematic decoupling of the cold
and ionized gas from the stars
\citep{Sarzi:2006p1474,2011MNRAS.417..882D,2014MNRAS.438.2798K}

In this paper we consider how gas accretion and depletion affect this
last piece of observational evidence: the kinematic decoupling of the
cold and ionized gas from the stars. In Section~\ref{sec:timescales}
we discuss how the observed distribution of angles between the
projected angular momenta of the gas and stars can be used to estimate
the gas-rich merger rate of ETGs and the time taken for gas to be
depleted. In Section~\ref{sec:models} we use a simple model to show
that there is tension between these estimates and other
observations. We discuss how to reconcile the two in
Section~\ref{sec:discussion} and conclude briefly in
Section~\ref{sec:conclusions}.
%
%
\section{ACCRETION AND DEPLETION TIMESCALES}
\label{sec:timescales}
%
%
\subsection{Origin of the gas}
\label{sec:origin}
\citet{2011MNRAS.417..882D} studied the kinematic misalignment (the
difference between the projected angular momenta) of the molecular gas
and stars in the complete, volume-limited ATLAS$^{\rm 3D}$ survey of
ETGs \citep{2011MNRAS.413..813C}. Following \cite{Sarzi:2006p1474},
they assumed that gas that has cooled from stellar mass loss must
always be kinematically aligned with those same stars. On the other
hand, material that has been accreted onto a galaxy from external
sources can have any angular momentum. \citet{2011MNRAS.417..882D}
found that roughly two-thirds of the galaxies are kinematically
aligned and are thus {\em consistent} with an internal gas origin
(i.e.\ stellar mass loss and/or gas remaining gas from the galaxy
formation). One-third of the galaxies are kinematically misaligned and
{\em must} therefore have acquired their gas externally (i.e.\ cold
accretion and/or minor mergers).

\begin{figure}
\begin{center}
\includegraphics[width=0.5\textwidth]{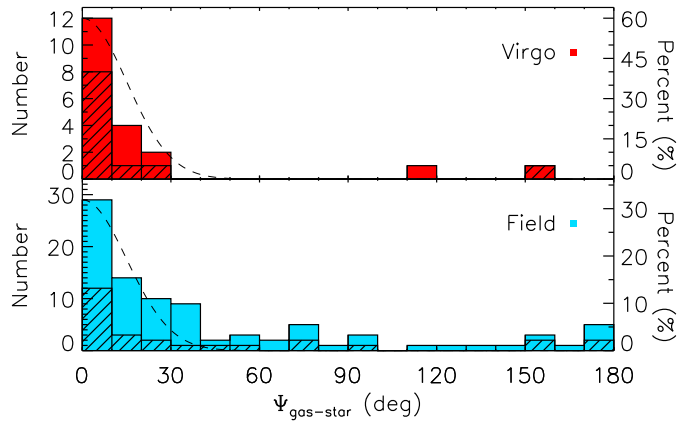}
\caption{Distribution of kinematic misalignment between gas and stars
  in the ETGs of the ATLAS$^{\rm 3D}$ sample. Ionised gas is shown in
  solid colours and molecular gas as the shaded histogram. {\bf Top:}
  Virgo cluster galaxies. {\bf Bottom
    panel:} Field and low-density environment galaxies.  Adapted from \citet{2011MNRAS.417..882D}.}
\label{fig:misalignment}
\end{center}
\end{figure}

Most importantly, as shown in Figure~\ref{fig:misalignment}, this
behaviour varies drastically as a function of environment. Indeed,
there is essentially {\em no} kinematically-misaligned galaxy in the
Virgo cluster (the only cluster surveyed by ATLAS$^{\rm 3D}$),
implying that external accretion of cold gas is shut off in dense
environments. Conversely, at least {\em half} of the galaxies in the
field have acquired their molecular gas externally. The same applies
to ionised gas probed through optical emission lines, {that can be detected in some cases in objects with ionised gas masses as low as $\approx10^3$ \msun}. This results in a higher
detection rate overall ($61\%$ for ionised gas versus $22\%$ for
molecular gas; \citealt{Sarzi:2006p1474,2011MNRAS.414..940Y}) and thus
better number statistics.
%
%
%
%
\subsection{Depletion timescale}
\label{sec:depletion}
Although its importance was not recognised at the time, a key feature
of the histograms presented in Figure~\ref{fig:misalignment} is the
absence of a peak at a kinematic misalignment of $180\degr$ for field
galaxies. A kinematic misalignment generally traces a morphological
misalignment \citep{2011MNRAS.414.2923K}, and structurally-misaligned gas in a fast-rotating ETG
(one with a clear disc structure) should be forced back to the
equatorial plane over time. This is primarily because the disc gravity
provides a substantial torque toward the plane, and because gas
dissipation prevents long lived precession of the gas reservoir
\citep{1982ApJ...252...92T}. To first order this relaxation should
occur symmetrically, with only the initial accretion angle determining
if the gas relaxes to be exactly co- or counter-rotating, although
stellar mass loss (co-rotating by definition) should create a drag on
any misaligned material (including that at $180\degr$). The peak of
galaxies with kinematically-aligned gas (at
$|\psi_{\rm gas-stars}|=0\degr$) is thus likely caused by a
combination of this relaxation process for externally accreted gas,
drag from stellar mass loss, and gas with an internal origin (i.e.\
cooled stellar mass loss itself). By contrast, the bin at $180\degr$
should contain only relaxed externally-accreted gas.
  
Assuming we are not observing local ETGs at a special epoch in their
evolution, the absence of a peak at a kinematic misalignment angle
$|\psi_{\rm gas-stars}|=180\degr$ in Figure~\ref{fig:misalignment}
thus implies that the externally-accreted gas is depleted on a
timescale $t_{\rm dep}$ (equal to the gas survival timescale
$t_{\rm surv}$) shorter than the torquing timescale $t_{\rm torque}$,
irrespective of the depletion process.

Previous studies of the relaxation of gas discs in
elliptical galaxies were conducted to understand the
dynamics of gaseous polar rings. \cite{1982ApJ...252...92T} and
\cite{1983ApJ...270...51L} considered relaxation in both axisymmetric
and triaxial halos, and found that in either case the alignment
process typically takes a few dynamical times ($t_{\rm
  dyn}$). \cite{1983ApJ...270...51L} found that
\begin{equation}
t_{\rm torque} \approx \frac{t_{\rm dyn}}{\epsilon},
\end{equation}
where $\epsilon$ is the ellipticity of the potential. For a typical
bulge in a lenticular galaxy $\epsilon\approx0.2$
\citep{2008A&A...478..353M}, thus
$t_{\rm torque}\approx5\,t_{\rm dyn}$.

As shown in \citet{2013MNRAS.429..534D}, the CO discs of the
ATLAS$^{\rm 3D}$ ETGs have typical radii $R\approx1$~kpc, with
rotational velocities at the disc edges of
$V_{\rm rot}\approx200$~km~s$^{-1}$. This therefore yields the
following constraint on the accreted molecular gas depletion
timescale:
\begin{eqnarray}
t_{\rm dep}\,=\,t_{\rm surv} & \leq & \,t_{\rm torque} \nonumber\\
                        & \la  & \frac{t_{\rm dyn}}{\epsilon} \nonumber\\
                        & \la  & \frac{2\pi\,R\,/\,V_{\rm rot}}{\epsilon} \nonumber\\
                        & \la  & \frac{3\times10^7\,{\rm yr}}{0.2} \nonumber\\
                        & \la  & 1.5\times10^8\,{\rm yr}\,\,\,.
\label{eq:tdep}
\end{eqnarray}

The estimate $t_{\rm dep}\,\la\,10^8$~yr is surprisingly
short. Indeed, if the accreted molecular gas were to form stars at the
same rate as observed in later-type disc galaxies (i.e.\ spiral
galaxies), it should be consumed on a timescale of
$1$-$2\times10^9$~yr
\citep[e.g.][]{1998ApJ...498..541K,2011ApJ...730L..13B}. Furthermore,
the star formation efficiency of normal ETGs is {\em lower} than that
of spiral galaxies
\citep[e.g.][]{2007ApJ...669..232K,2013MNRAS.432.1914M,2012ApJ...758...73S,2014MNRAS.444.3427D},
so if anything the depletion timescale of molecular gas through star
formation should be even longer in ETGs. According to the arguments
above, the accreted gas observed in local ETGs must therefore be
depleted at least $10$ times faster than the star-formation
consumption timescale. We discuss this puzzling result in more depth
in Section~\ref{sec:discussion}.
%
%
\subsection{Accretion timescale}
\label{sec:accretion}
Combining the accreted molecular gas depletion (or survival) timescale
with our detection rate ($N_{\rm gas}$/$N_{\rm tot}$) yields a simple
estimate of the accretion timescale $t_{\rm acc}$, that is the time
interval between successive accretion events:
$t_{\rm acc}\,\approx\,t_{\rm dep}\,/(N_{\rm gas}/N_{\rm tot})$.

The overall molecular gas detection rate of $22\%$
\citep{2011MNRAS.414..940Y} is an overestimate of the true value, as
it includes kinematically-aligned galaxies to which the current
argument does not apply. Only kinematically-misaligned gas should be
counted. About one-third of all the detected galaxies are
kinematically-misaligned, but this is an equally misleading
(under)estimate, since we know that accretion does not take place in
dense environments \citep{2011MNRAS.417..882D}. Indeed, the detection
rate of kinematically-misaligned gas in the Virgo cluster is
essentially nil, effectively yielding an infinite gas accretion
timescale. The right detection rate to use is therefore that of
kinematically-misaligned molecular gas in the field (and low-density
environments such as groups). The overall molecular gas detection rate
in the field is $29\%$ \citep{2011MNRAS.414..940Y}, and roughly half
of this gas is kinematically-misaligned \citep{2011MNRAS.417..882D},
yielding a relevant detection rate of $\approx14\%$. The molecular gas
accretion timescale is therefore
\begin{eqnarray}
t_{\rm acc} & =  & \frac{t_{\rm dep}}{(\frac{N_{\rm gas}}{N_{\rm tot}})|_{\rm field}} \nonumber\\
         & \la & \frac{t_{\rm dyn}\,/\,\epsilon}{(\frac{N_{\rm gas}}{N_{\rm tot}})|_{\rm field}} \nonumber\\
         & \la & \frac{1.5\times10^8\,{\rm yr}}{0.14} \nonumber\\
         & \la & 10^9\,{\rm yr}\,\,\,.
\label{eq:tacc_mol}
\end{eqnarray}

Our estimate therefore suggests that local ETGs in the field accrete
molecular gas from an external source at least once every gigayear
($R_{\rm acc}\gtsimeq1$~Gyr$^{-1}$). We discuss this estimate further,
and compare it with those of other studies, in
Section~\ref{sec:discussion}.

 Interestingly, the overall detection rate of diffuse ionised gas emission (H$\beta$ and/or [OIII] emission lines) in the ATLAS$^{\rm 3D}$ sample is
  $61\%$ (Sarzi et al., in prep). As \citet{2011MNRAS.417..882D} showed that the
 ionised and molecular gas are always kinematically-aligned and
 therefore must share a common origin, we can in principle replace the molecular
 gas detection rate used in Equation~\ref{eq:tacc_mol} by that for
 the ionised gas. The ionised gas detection rate for field galaxies
 in ATLAS$^{\rm 3D}$ is $68\%$, with roughly half
 kinematically-misaligned, so the relevant detection rate is
 $34\%$, roughly $2.5$ times higher than that used above for the molecular gas. This in turn yields an upper limit on the ionised gas accretion
 timescale for local field ETGs that is $2.5$ times smaller, only $\la\,4\times10^8$~yr.

 The relation between the gas phases in objects without detectable cold gas is, however, unclear. In many objects the ionised gas is likely in an interface layer (ionised by old stars, AGN or low level star-formation) around a atomic/molecular gas disc that has a low enough mass to fall below our sensitivity limit. This interpretation is supported by the similarity in the misalignment histograms (Figure \ref{fig:misalignment}). It is possible that this emission could be more nebulous, however, arising from the envelopes around evolved stars, and thus not be in a disc at all in some objects (in this case, however, the ionised gas emission would have to co-rotate with the stars).
Given the above uncertainty the estimates of \citet{1982ApJ...252...92T} and
\citet{1983ApJ...270...51L} may not apply (or, equivalently, the scaling factor may be different), and the ionised gas relaxation time may be much longer than that for molecular gas. This would then yield larger estimates for both $t_{\rm dep}$ and $t_{\rm acc}$. Given these uncertainties, in the rest of this paper we favour the more reliable molecular gas estimates to put other timescales in context.
%
%
\section{ACCRETION MODELS}
\label{sec:models}
%
%
\subsection{Toy  model}
\label{sec:mod_toy}
To validate and explore further the simple calculations above, we
created a toy model. This model includes in a simple framework the
processes of gas accretion, relaxation and depletion. These are
characterised by three free parameters, respectively the accretion
rate ($R_{\rm acc}$, the number of gas accretion events per unit
time), the relaxation timescale ($t_{\rm relax}$; in units of
$t_{\rm dyn}$), and the gas depletion time ($t_{\rm dep}$).

We initialised the model at time $t=0$ with an arbitrary number of
gas-free, ``red and dead'' ETGs. These have realistic luminosities
drawn uniformly from a Schechter function representation of the
ATLAS$^{\rm 3D}$ $K$-band galaxy luminosity function
\citep{2011MNRAS.413..813C}. From this we calculate the maximum
circular velocity at which the gas can rotate by assuming the galaxies
follow the $K$-band CO Tully-Fisher relation of
\cite{2011MNRAS.414..968D}. Molecular gas reaches beyond the turnover
of the rotation curve in $\approx70\%$ of ATLAS$^{\rm 3D}$ ETGs
\citep{2011MNRAS.414..968D,2013MNRAS.429..534D}, so this should
provide a good estimate of the rotation velocity of the gas at the
edges of the gas discs for the bulk of the population.

After each timestep we randomly select galaxies that will accrete gas,
such that the average number of mergers per galaxy per unit time is
equal to $R_{\rm acc}$. {In this simple model, $R_{\rm acc}$ is independent of galaxy mass and morphology, which may not be true in reality. We discuss this assumption further in Section \ref{discussuncert}.} 

\begin{figure}
  \begin{center}
    \includegraphics[width=0.45\textwidth,angle=0,clip,trim=0.0cm 0cm 0cm 0.0cm]{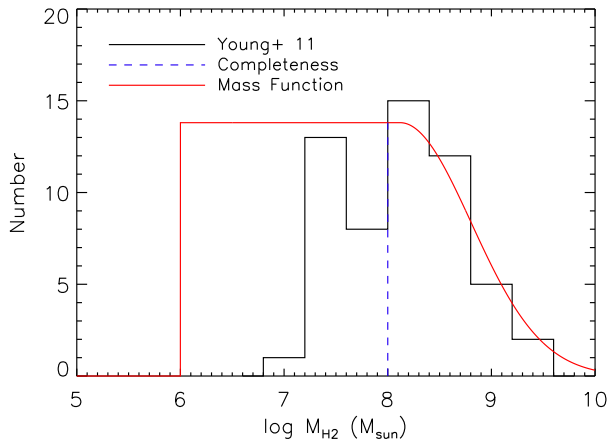}
  \end{center}
  \caption{Observed and modeled H$_2$ mass functions. The measured
    distribution of H$_2$ masses from
    \protect\cite{2011MNRAS.414..940Y} is shown in black, with the
    approximate completeness limit indicated by a dashed line. The
    (arbitrarily normalised) H$_2$ mass function used in the toy
    models is shown in red (see Section~\ref{sec:mod_toy}).}
  \label{fig:h2massfunc}
\end{figure}

The amount of gas accreted is chosen randomly
from the H$_2$ mass function of \cite{2011MNRAS.414..940Y}, that was
fitted with a Gaussian at the high mass end, and was assumed constant
below the completeness limit down to an H$_2$ mass of $10^6$~\msun\
 (see Fig.~\ref{fig:h2massfunc}), as predicted from semi-analytic
models \citep[e.g.][]{2014MNRAS.443.1002L}. Removing the low-mass
cutoff does not affect our results. 
{We here assume that after coalescence each one of our mergers results in the creation of a disc of H$_2$ in the centre of the remnant, with a mass that is drawn from the observed distribution. For this simple model, we condense the complex time-dependent processes happening in mergers, and assume that this final state is obtained instantaneously. We discuss these assumption further in Section \ref{discussuncert}.}

We determine the radial extent of the gas, and hence the dynamical
time at its outer edge, by drawing uniformly from a log-normal fit to
the observed distribution of molecular disc radii in
\cite{2013MNRAS.429..534D}. 

The angle at which gas is accreted onto galaxies is also chosen
uniformly, sampling the full range $0$-$180^{\circ}$. If gas already
exists in a galaxy, the interaction between the two reservoirs is
roughly included by mass-weighting the resulting position angle of the
gas (thus lying between the newly accreted and old material). {Our assumption about the exact form of this weighting does not strongly affect our results (see Section \ref{discussuncert}).}
 
%
%

The relaxation of the gas is tracked at each timestep. The gas relaxes
at a rate that has a cosine dependance on the current misalignment
angle (i.e.\ the rate increases as the disk approaches the galaxy
plane, where the torque is higher), following the formalism of
\cite{1982ApJ...252...92T}. The total time taken to relax is
calculated such that the gas disc relaxes into the plane from a
misalignment of $60\degr$ in a time $t_{\rm relax}$. We note that
adopting a simpler form, where the gas relaxes linearly (i.e.\ by a
full $\pi/2$ radians in a time $t_{\rm relax}$) does not affect the
results of this paper. More complex prescriptions for the gas
relaxation are beyond the scope of our analysis, but we expect any
reasonable form to not unduly alter the results derived from these two
cases.

Finally, we destroy a fraction of the molecular gas at each time
step. In the first set of simulations discussed below, we choose to
include only the effect of star formation on the molecular gas, using
up $\approx10\%$ of the gas per dynamical time, equivalent to the
standard star-formation law derived from local star-forming galaxies
\citep{1998ApJ...498..541K}. In Section~\ref{sec:agn}, we also include
a very simple model of gas destruction (such as that caused by AGN
feedback). An additional free parameter ($t_{\rm AGN}$) then sets the
duty cycle over which gas is destroyed. All the gas is instantaneously
blown out of the model galaxy a time $t_{\rm AGN}$ after it has been
accreted. Clearly, this highly simple model is not meant to capture in
detail the highly complex AGN feedback process, but it at least allows
us to explore in a simple manner its effect. Any process that removes
gas in one discrete event can be modelled by this formalism.

At the end of each timestep, we are thus left with the observable gas
mass and kinematic position angle of each galaxy. We convolve the
kinematic misalignment angles with a Gaussian kernel to match the
observational errors, and set a limit of $10^7$~\msun\ to define
objects that are detectable, this to match the observations of the
ATLAS$^{\rm 3D}$ survey (see Fig.~\ref{fig:misalignment};
\citealt{2011MNRAS.414..940Y}). Each model presented in
Section~\ref{sec:mod_results} includes $10^5$ simulated galaxies and was run for $5$~Gyr, although {the misalignment distribution histograms reach} a steady state more quickly.
%
%
\subsection{Model results}
\label{sec:mod_results}

\subsubsection{Reference model}
\label{sec:reference_run}
\begin{figure}
  \begin{center}
    \includegraphics[width=0.45\textwidth,angle=0,clip,trim=0.0cm 0cm 0cm 0.0cm]{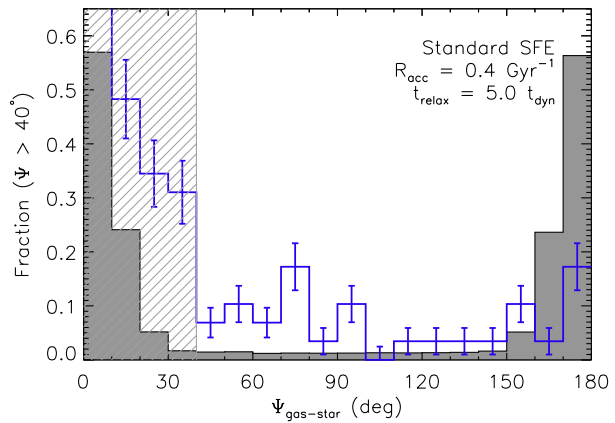}
  \end{center}
  \caption{Kinematic misalignment histogram of our reference model,
    showing the steady state kinematic misalignment distribution (grey
    histogram) of a model where the gas is depleted only by
    star-formation with a standard efficiency, and where the gas
    relaxes in $5$ dynamical times. The accretion rate required to
    match the observed detection rate is then $0.2$~Gyr$^{-1}$. The
    parameters used are listed in the figure legend. The
    ATLAS$^{\rm 3D}$ data are shown as a blue histogram, and should
    not be compared with the model in the shaded region below
    $40\degr$, where other processes that are not modelled here may
    dominate (e.g.\ stellar mass loss), thus increasing the number of
    detected objects. Both histograms are normalised so the area with $\Psi>40\degr$ is equal to unity.}
  \label{fig:reference_run}
\end{figure}

In Figure~\ref{fig:reference_run}, we present the misalignment
histogram resulting from our reference model, based on our best
current understanding of the processes involved. This model uses only
the standard star-formation law to deplete the gas, a relaxation time
of $5\,t_{\rm dyn}$ (as derived from the results of
\citealt{1983ApJ...270...51L}), and an accretion rate of
$0.4$~Gyr$^{-1}$ (calculated so the detection rate of
kinematically-misaligned objects in the model matches the observed
rate). This accretion rate is consistent with observations of the
(minor and minor+major) merger rate estimated from other methods
(see Section \ref{sec:discuss_gaslife}).

%
%

{One should not compare the misalignment histogram produced by the model to the data within the shaded region in Figure~\ref{fig:reference_run} ($\Psi<40\degr$), as material produced by internal processes not included in our model (e.g. stellar mass loss) will contribute (and may dominate) here. Outside this region,} clear discrepancies are immediately obvious when comparing the
reference model histogram to the observed one
(Fig.~\ref{fig:reference_run}). First, almost all model galaxies
($\approx90\%$) lie in the exactly kinematically aligned and
misaligned bins (at $0$ and $180^{\circ}$, respectively). Second, the
bins for kinematically aligned and misaligned gas contain
approximately equal numbers of model galaxies. To produce an
asymmetric histogram, one therefore clearly requires another process
that preferentially produces/preserves gas that is always kinematically aligned
with the stars (or, conversely, preferentially removes gas that is
kinematically misaligned). The reference model thus highlights
graphically the points made in Section~\ref{sec:depletion}, that the
standard relaxation and gas depletion timescales cannot reproduce the
observed misalignment histograms. In the sections below, we thus
consider possible solutions to this problem.
\subsubsection{Quicker gas depletion}
\label{sec:tdep_run}
\begin{figure}
  \begin{center}
    \includegraphics[width=0.45\textwidth,angle=0,clip,trim=0.0cm 1.6cm 0cm 0.0cm]{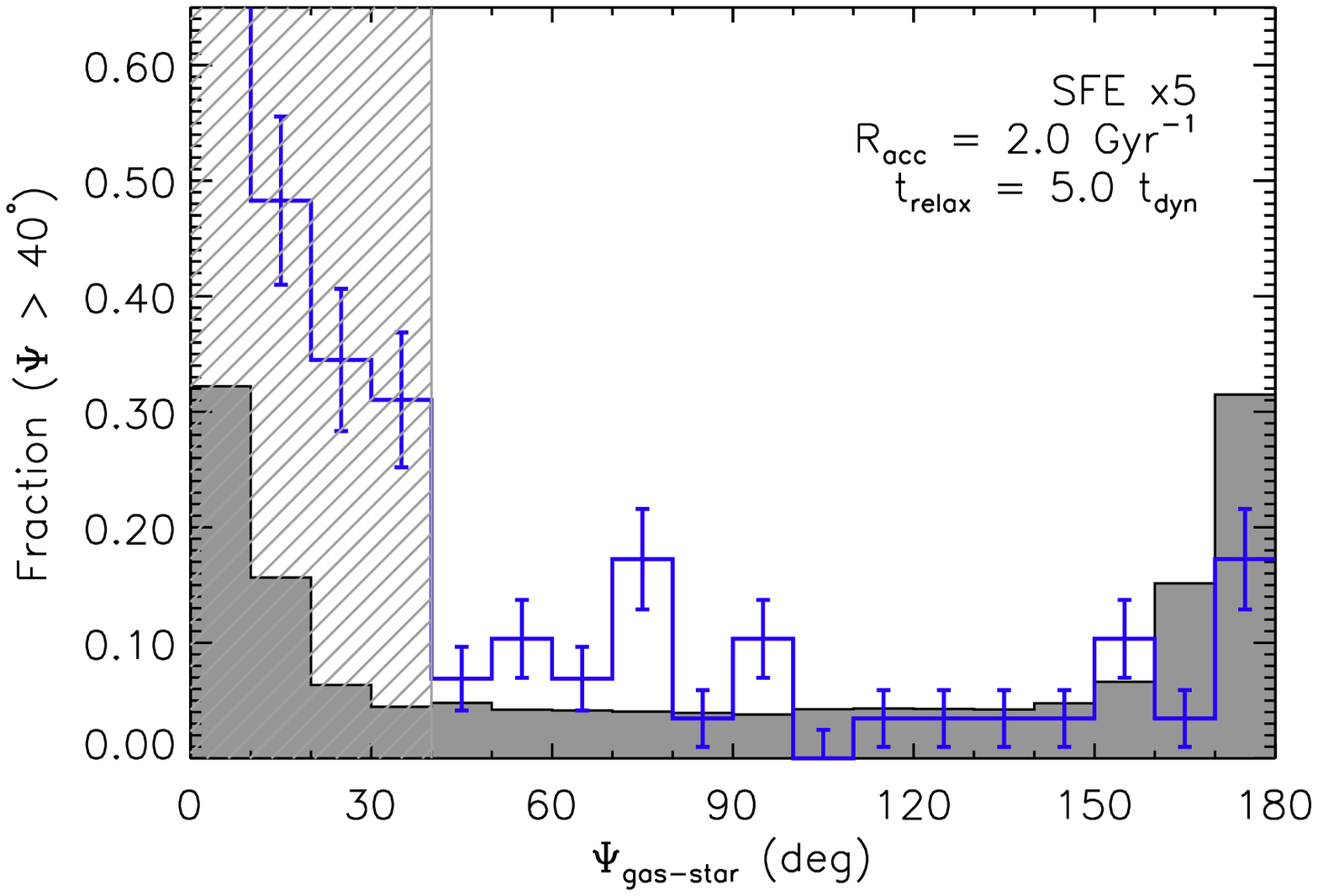}
    \includegraphics[width=0.45\textwidth,angle=0,clip,trim=0.0cm 0cm 0cm 0.0cm]{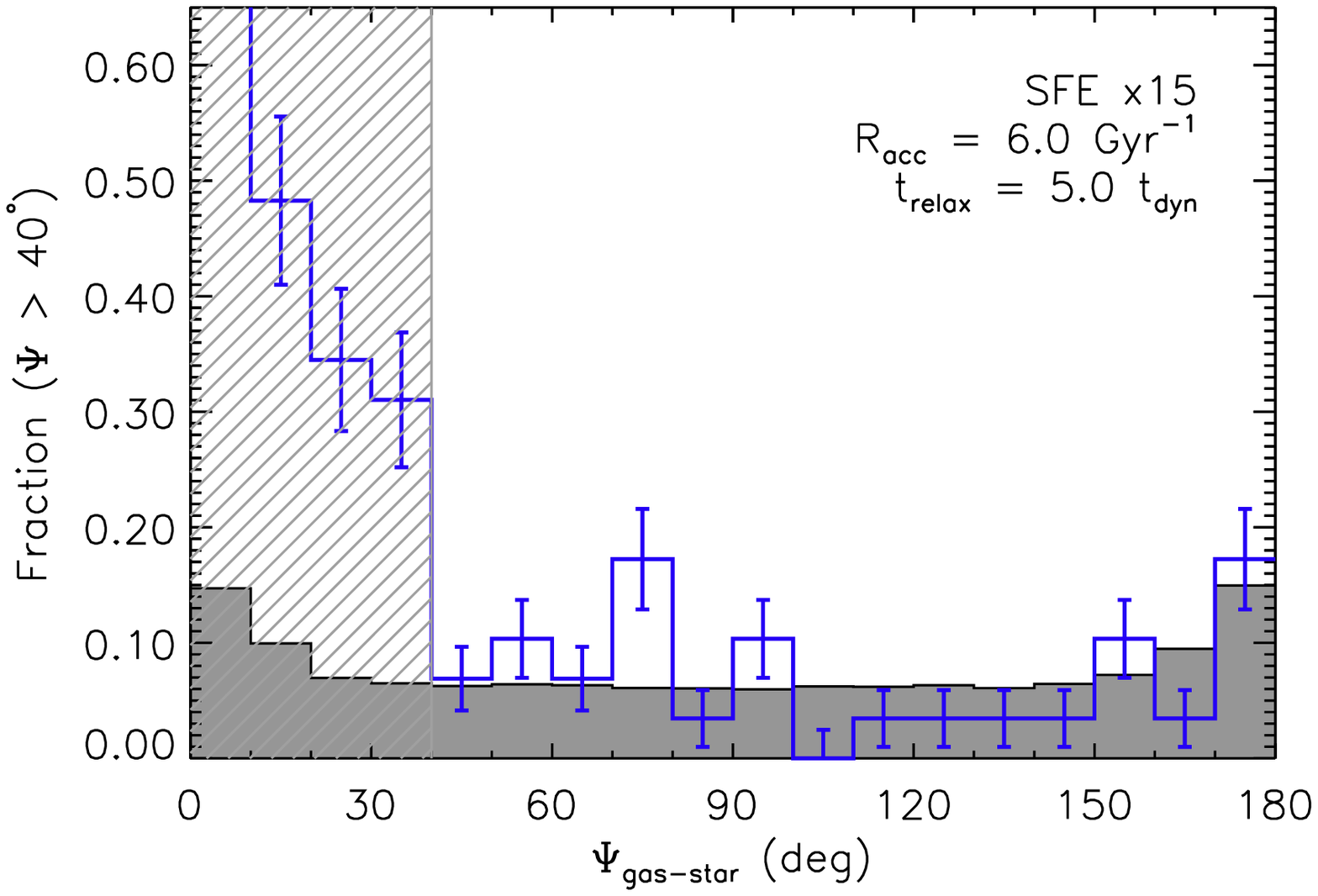}
  \end{center}
  \caption{As Figure~\ref{fig:reference_run}, but for an enhanced
    star-formation efficiency of $5$ (top panel) and $15$ (bottom
    panel) times the standard value. The accretion rates required to
    match the observed detection rate are then $2$ and
    $6$~Gyr$^{-1}$, respectively.}
  \label{fig:tdep_run}
\end{figure}

As argued in Section~\ref{sec:depletion}, a faster gas depletion time
may be required to understand the observed gas kinematic misalignment
histograms. We show in Figure~\ref{fig:tdep_run} a model where the
star-formation efficiency (SFE) has been increased by a factor of $5$
(top panel) and $15$ (bottom panel) with respect to its standard
value. This increase can be used to model any process that removes gas
over time, not just star formation. Clearly, an order of magnitude
increase in the depletion rate is required, over and above that
expected from normal star formation, before the distribution of
misalignments in the model can match that of the observations.

%
%

One drawback of this solution is that the accretion rate required to
match the observed detection rate is inversely proportional to the gas
depletion time, that is shorter depletion times require proportionally
more accretion events per unit time to maintain the same detection
rate. In these models with increased SFEs, one thus requires a merger
rate of $\approx6$~Gyr$^{-1}$ to match the observed detection rate. We
discuss this hypothesis further in Section~\ref{sec:discuss_gaslife}.
\subsubsection{Gas destruction}
\label{sec:agn}
In Section~\ref{sec:depletion}, we discussed the need to decrease the
survival time of gas in ETGs. One mechanism that has become popular in
recent years to quench ETGs is AGN feedback
\citep[e.g.][]{1998A&A...331L...1S,2012ARA&A..50..455F}. In this
paradigm, the central black hole injects a significant amount of
energy into the ISM, removing or destroying it. As described in
Section~\ref{sec:mod_toy}, we can include a simple feedback
prescription in our model, that blows all of the gas out of the
galaxies a time $t_{\rm AGN}$ after the gas has been accreted.

\begin{figure}
  \begin{center}
    \includegraphics[width=0.45\textwidth,angle=0,clip,trim=0.0cm 0cm 0cm 0.0cm]{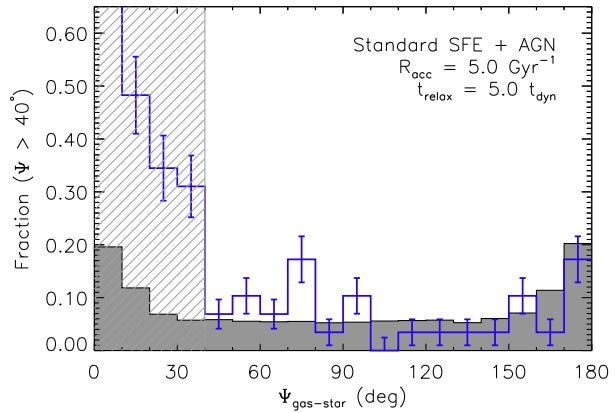}
  \end{center}
  \caption{Same as Figure~\ref{fig:reference_run}, but with our simple
    gas destruction prescription enabled (approximating AGN
    feedback). The assumed AGN duty cycle is $5\times10^7$~yr. The
    accretion rate required to match the observed detection rate is
    then $5$~Gyr$^{-1}$.}
  \label{fig:agn_run}
\end{figure}

We show the result of turning on this simple AGN feedback prescription
in Figure~\ref{fig:agn_run}. We use the same standard SFE and
relaxation timescale as the reference model
(Section~\ref{sec:reference_run}), and additionally set
$t_{\rm AGN}=10^8$~yr. This is at the upper end of the range
$10^6$-$10^8$~yr that has been suggested in the literature for the
typical AGN duty cycle
\citep[e.g.][]{2001ApJ...549..832S,2002ApJ...567L..37M,2006ApJS..163....1H}. We
again adjust the accretion rate to match the observed detection rate
of molecular gas. The AGN feedback prescription does lower the
fraction of galaxies within the co- and counter-rotating bins
sufficiently to match the observed predictions. However, it also
requires a very high accretion rate (approximately $10$ times
higher than required in the reference model) to match the
observed CO detection rate of kinematically-misaligned objects.

\subsubsection{Slower relaxation}
\label{sec:trelax_run}
\begin{figure}
  \begin{center}
    \includegraphics[width=0.45\textwidth,angle=0,clip,trim=0.0cm 1.6cm 0cm 0.0cm]{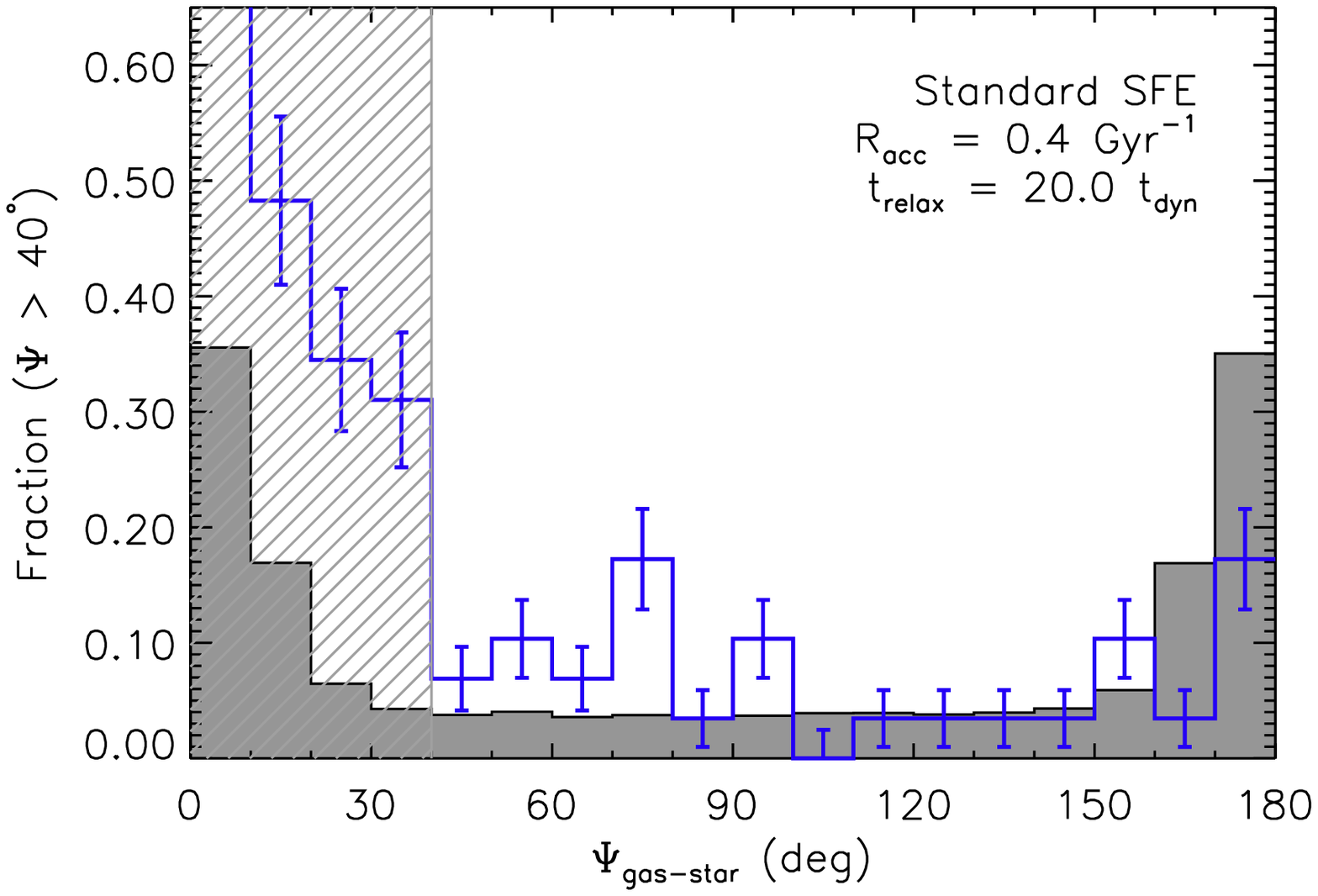}
    \includegraphics[width=0.45\textwidth,angle=0,clip,trim=0.0cm 0cm 0cm 0.0cm]{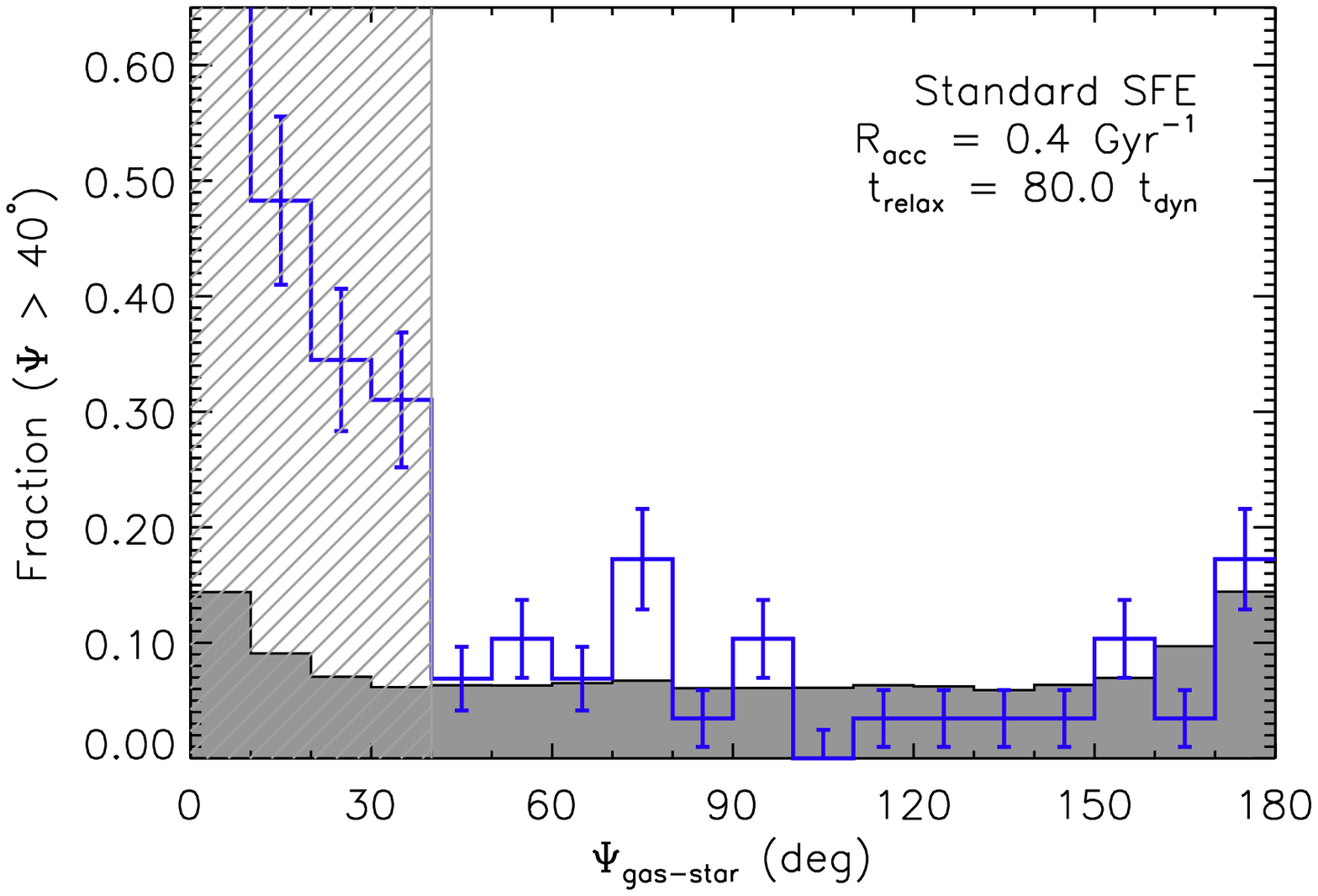}
  \end{center}
  \caption{Same as Figure~\ref{fig:reference_run}, but for a slower
    gas relaxation of $20$ (top panel) and $150$ (bottom panel)
    dynamical times.}
  \label{fig:trelax_run}
\end{figure}

An alternative explanation for the large fraction of model galaxies
whose gas reservoirs have yet to relax is that this process is slower
than usually assumed. Figure~\ref{fig:trelax_run} shows model runs
assuming a standard SFE for depleting the gas and a standard accretion
rate of $0.4$~Gyr$^{-1}$ (so the detection rate in the model matches
the observed rate by construction, as in
Section~\ref{sec:reference_run}). The relaxation time is however
increased to $20$ and $80$ $t_{\rm dyn}$, respectively, in the top
and bottom panels. These models show that the relaxation time would
have to be much longer than normally assumed (by around an order of
magnitude) to reproduce the observed lack of counter-rotating
objects. This solution has the benefit that it removes the need for an
extremely large accretion rate, but it requires additional processes
to slow gas relaxation. We discuss this further in
Section~\ref{sec:discuss_slowrelax}.
%
%
%
%
\section{Discussion}
\label{sec:discussion}
In the sections above, we discussed the distribution of kinematic
misalignments between the gas and stars of local ETGs, and the way in
which it encodes information about the depletion and accretion
timescales of the gas. We showed that the standard gas relaxation and
depletion timescales from the literature cannot reproduce the observed
misalignment histogram. We discuss below possible solutions to this
problem.
\subsection{Gas depletion and destruction}
\label{sec:discuss_gaslife}
In Section~\ref{sec:tdep_run}, we showed that one can reproduce the
observed kinematic misalignment histograms from ATLAS$^{\rm 3D}$ by
increasing the rate at which gas is consumed. According to this
argument, the accreted gas observed in local ETGs would have to be
depleted at least $10$ times faster than the standard star-formation
consumption timescale. Extreme starbursts have been observed to have
such small depletion times \citep[e.g.][]{1994ApJ...431L...9M}, but
studies of ETGs
\citep[e.g.][]{2012ApJ...758...73S,2013MNRAS.432.1914M,2014MNRAS.444.3427D}
show that their SFE is actually slightly lower than in spiral galaxies
(and therefore very much lower than in starburst objects). Fast
starbursts thus cannot be invoked to explain the short depletion times
required here. Other potential physical mechanisms that could remove
gas include dissociation of H$_2$ by X-rays from the ETGs' hot gas
halos, friction and/or shocks against kinematically-aligned gas from
stellar mass loss of the bulk (co-rotating) galaxy stellar
populations, and AGN feedback. Whatever the depletion mechanism is, it
needs to be very rapid.

The depletion timescale of $\la10^8$~yr we estimated in Section
\ref{sec:depletion} is the same as the typical timescale over which
central supermassive black holes are active
\citep[e.g.][]{2001ApJ...549..832S,2002ApJ...567L..37M,2006ApJS..163....1H}.
The energy input from an AGN is observed to be expelling the gas from
at least some ETGs (e.g.\ NGC1266;
\citealt{2011ApJ...735...88A,2012MNRAS.426.1574D}), hence we tried to
include them in our toy models in a simple way (see
Section~\ref{sec:agn}). We then find that a duty cycle of $\la10^8$~yr
is sufficient to bring the model kinematic misalignments in line with
the observed data.

The drawback of both fast depletion and/or destruction of gas is that
the accretion rate required to match the observed detection rate is
inversely proportional to the gas depletion time. With such small gas
depletion/destruction timescales, a merger rate of $\approx5$~Gyr$^{-1}$ is required to maintain the detection rate
as observed. These timescales can also be used to constrain the likely
mechanism(s) supplying the gas. They rule out major mergers (that are
too rare), but as the accretion timescale becomes ever smaller, the case for 
nearly continuous accretion becomes stronger, and accretion of diffuse gas
(rather than galaxies) along filaments may need to be considered.
However, the large fraction of gas-rich ETGs that are
morphologically disturbed in deep imaging
\citep[e.g.][]{2005AJ....130.2647V,2014arXiv1410.0981D} suggests that
stars are also accreted along with the gas, leading to the idea that
minor mergers may dominate over cold mode accretion \citep[see
e.g.][]{2009MNRAS.394.1713K,2011MNRAS.411.2148K}.

Interestingly, there is evidence of two separate recent minor mergers
within such a short time interval in at least one ATLAS$^{\rm 3D}$ ETG
galaxy, NGC4150. Indeed, Kaviraj et al.\ (in preparation) have shown
that although this galaxy has a young ($\approx10^8$~yr)
counter-rotating stellar core, presumably formed in a gas-rich minor
merger, the molecular gas currently present is co-rotating with the
bulk of the stars, suggesting an even more recent accretion event.

In the current $\Lambda$CDM cosmology, the universe is (primarily)
hierarchical in nature, making galaxy mergers an essential aspect of
galaxy formation and evolution. The gas-rich merger rate at $z=0$ is
thus an interesting quantity to determine from the gas misalignment
data, and to compare with that measured from other observations and
simulations
\citep[e.g.][]{1993MNRAS.262..627L,2006ApJ...647..763M,2008MNRAS.384....2G,Genel:2009gt,2011ApJ...742..103L,2012ApJ...747...34B,2012ApJ...746..162N,2014MNRAS.445.1157C,2015arXiv150201339R}.

The rate of gas-rich mergers derived here is somewhat higher than
previous estimates of the incidence of minor mergers. For instance,
\cite{2011ApJ...742..103L} presented a range of estimates, and
compared it extensively with other samples. They estimated a total merger
rate of $0.2$ to $0.5$~Gyr$^{-1}$ with $75\%$ of these mergers being
minor (mass ratios of 1:4 to 1:10).
Theoretically, studies such as that of \citet{2010ApJ...715..202H}
find a major+minor merger rate of $\approx0.2$~Gyr$^{-1}$ for objects
of $10^{11}$~\msun, but predict that this merger rate scales
positively with galaxy mass. The more massive galaxies in our sample
(with stellar masses above $10^{12}$~\msun) are predicted to have
merger rates of up to $0.9$~Gyr$^{-1}$, still below our estimate. To
make things worse, the estimates discussed above count all mergers,
not only those that are gas rich.

We thus conclude that although accretion rates as fast as those
require here may be possible for individual objects, it is unlikely
that the average accretion rate in the universe is as high as our
models require.
\subsection{Slow gas relaxation}
\label{sec:discuss_slowrelax}
Another way to reconcile the observed kinematic misalignments of ETGs
with the theoretical predictions is to adopt a relaxation (i.e.\
torquing) timescale much longer than a few dynamical times. This
explanation is appealing, as it does not require high gas accretion
rates, but the possible physical processes slowing the relaxation down
are currently unconstrained. One possibility is that the accreted
misaligned gas simply precesses. This would be possible if the
accreted molecular gas were moving purely ballistically, as discrete
molecular clouds can do, but it seems unlikely for the complex
multi-phase ISM characteristic of galaxies, and especially for the
warmer more diffuse ionised gas.

If gas is accreted onto ETGs over long timescales, then the continued
addition of misaligned material could slow the relaxation
process. This could happen, for instance, if a gas-rich merger throws
out long tidal tails of gas, that fall back slowly over time
\citep[e.g.][]{Barnes:2002ke}. Atomic gas tails are observed in a few
objects \citep[e.g.][]{2006MNRAS.371..157M,2012MNRAS.422.1835S}, but
not in the entire population of misaligned ETGs. Low-density tails, or
accretion of debris in hotter gas phases, could however be missed. 
\cite{2015MNRAS.451.3269V} looked at this process in a hydrodynamic simulation of a massive early-type galaxy. 
They found that continued accretion did indeed slow relaxation, as the angular momentum change induced by accreted gas dominates over that induced by stellar torques. The total time taken for the disc to settle in these simulations was 80-100 $t_{\rm dyn}$, very similar to the timescale we estimate in this work. 

Another option is that the hot halos around galaxies have an
effect. \citet{2014MNRAS.443.1002L,2014arXiv1410.5437L} found that the
hot halos around ETGs may be misaligned (either because of mergers or
continuous accretion from the cosmic web), and that this could be the
cause of the observed kinematic misalignment distribution. These
models make various assumptions that still need to be verified through
hydrodynamic simulations, and they did not include any relaxation
process, but it is clear that continued misaligned cooling could
extend relaxation times. In addition, if the hot ISM of ETGs has a net
angular momentum and can exert pressure on the relaxing accreted gas,
then gas relaxation would not be a symmetric process, which would help
explain the observed asymmetric kinematic misalignment distribution.
These solutions, however, require the hot halos of galaxies to have
significant angular momenta. It will not be possible to
observationally verify whether the hot halos of galaxies have any
significant rotation, and/or if it correlates with the rotation of the
stellar body, until the X-ray satellite \textit{ASTRO-H} launches
\citep[see e.g.][]{2013MNRAS.434.1565B}.

\subsection{Model limitations and associated uncertainties}
\label{discussuncert}

{The toy model we present in this paper makes various assumptions, that could potentially affect the results we derive, and are thus discussed below.}

{We know from previous studies that the gas is not sufficiently extended to reach the flat part of the rotation curve in $\approx30\%$ of ETGs \citep{2011MNRAS.414..968D,2013MNRAS.429..534D}. In these cases our model will overestimate the gas rotation velocity, and thus underestimate the dynamical time, relaxing the gas too quickly. However, this always happens in gas reservoirs with small radial extents, and thus the shortest dynamical times. Even an underestimation of the dynamical time by a factor of two in these objects (the largest difference between the CO velocity and the circular velocity in objects with small gas reservoirs; \citealt{2011MNRAS.414..968D}) would not help reconcile the reference model and the observations. }

{The second assumption we make is that $R_{\rm acc}$, and the amount of gas accreted in each episode, is independent of galaxy mass and morphology. While this may be a reasonable first order approximation, it is known that ETG's follow a morphology-density relation \citep[e.g.][]{1980ApJ...236..351D}, and interactions preferably happen in group scale environments. We minimise the effect of any such bias here by only comparing the model to observed galaxies in field environments. In addition, the typical merger mass-ratio must vary as a function of galaxy mass, which will affect the typical amount of gas brought in by each encounter. However, as the typical gas fraction in observed ETGs does not seem to depend on galaxy mass or environment \citep{2011MNRAS.414..940Y} it is likely that this second effect is small, and would not change our results significantly. }

{Our approach assumes that all of the gas brought in by the merger is molecular. We do not model additional \hi\ gas, that may be brought in by the merger and settle in a larger scale disc (e.g. \citealt{2012MNRAS.422.1835S}), and could convert to H$_2$ over longer timescales. If significant radial mass/angular momentum exchange occurs, these larger scale \hi\ discs could slow relaxation similarly to the atomic gas tidal tails discussed above. }

{When gas is already present in a galaxy that experiences a new accretion event, we mass weight the misalignment angle of the new and old gas components to determine the resultant kinematic misalignment. In a real gas-rich merger complex processes take place that we are unable to capture in such a simple model, but we do not expect this simplification to have a major effect. Within our model, when this situation occurs the median amount of accreted gas is an order of magnitude larger than the gas already present, and thus the correction typically changes the accreted angle by only a small amount (a median change of $\approx4^{\circ}$). Ignoring this correction entirely, and instead forcing all the gas to rotate with the newly accreted material does not change our result.}

{We also assume that the radial extent of the gas in these objects is uncorrelated with its mass (and other galaxy properties). We expect this assumption to not unduly affect our results, as although a strong relation between the gas mass and size of atomic gas discs exists \citep[e.g.][]{1997A&A...324..877B}, any such relation for molecular gas is weak \citep{2013MNRAS.429..534D}. We do, however, caution that when gas is accreted onto a galaxy with an existing cold gas reservoir the interaction between the two discs could systematically affect the final radial extent of the settled gas disc.}

{In our simple model the SFE is a free parameter, but it does not depend on galaxy properties. Gas-rich mergers in the real universe seem to have increased SFEs, that result in faster gas depletion. These events are short-lived, however, and observed ETGs are found to have low star-formation efficiencies \citep[e.g.][]{2011MNRAS.415...61S,2014MNRAS.444.3427D}. Evidence also exists that gas minor mergers with early-type galaxies can further suppress the SFE \citep{2015MNRAS.449.3503D}. We thus do not expect merger induced starbursts to be important in these sources.}

 {Overall the limitations discussed above are mostly expected to be second-order effects. Hydrodynamic simulations simulations are required to confirm this, and they will be the topic of a future work.}

%
%
\section{CONCLUSIONS}
\label{sec:conclusions}
In this paper we considered what can be learnt about the processes of
gas accretion and depletion from the observed distribution of
kinematic misalignment angles between the cold/ionized gas and the
stars.

We first presented simple arguments showing that the misalignment
distribution encodes information on the relaxation, depletion and
accretion timescales of gas in ETGs. Specifically, we argued that the
lack of a peak of exactly counter-rotating objects strongly constrains
these timescales. Simple calculations, based only on a rough estimate
of the relaxation (i.e.\ torquing) timescale, imply a short gas
depletion timescale ($t_{\rm dep}\,\la\,10^8$~yr) and a high rate of
gas-rich mergers ($\ga\,1$~Gyr$^{-1}$).

We then presented a toy model of the interplay between these
processes, allowing us to better constrain and explore the effects of
these timescales on the kinematic misalignment distribution. Realistic
distributions of stellar masses and gas radii allow to derive
estimates of the relevant timescales for the whole ETG population. We
thus confirmed our simple calculations, clearly showing that the
standard values for the accretion rate, star formation efficiency and
relaxation time are not simultaneously consistent with the observed
distribution of kinematic misalignments.

We then explored the effect of varying these parameters. Both faster
gas depletion (via e.g.\ more efficient star formation) and/or faster
gas destruction (via e.g.\ AGN feedback) can be used to explain the
kinematic misalignment distribution, but they then require high rates
of gas-rich mergers ($\approx5$~Gyr$^{-1}$). Although some objects have
evidence of multiple mergers within such a timescale, as a population
accretion rates this high are unlikely.

An alternative explanation, which does not require high accretion
rates, is that the misaligned gas relaxation occurs over longer
timescales ($\simeq100$ $t_{\rm dyn}$ or $\approx1$-$5$~Gyr) than
usually assumed. We suggest that this could come about because of
ongoing accretion of tidal debris in mergers, rotating hot gas halos,
or continued accretion from the cosmic web.

One way in which these scenarios could be tested would be to determine
the age of the accretion event in other ways, such as age dating the most recent burst of star formation (via stellar
population modelling), or with the ratio of gas phase metallicity to stellar metallicity. Models with fast relaxation require the
misaligned objects to have accreted gas recently (typically within the
last $200$~Myr, but up to $\approx1$~Gyr for exactly counter-rotating
gas). Slow relaxation removes such differences, predicting a flat
distribution of accretion times as a function of misalignment. 

Of course, all of the processes discussed above may take place
simultaneously in galaxies and can not easily be separated. Many
combinations of the parameters in our models can match the observed
misalignment distribution. Further work, both observationally and
numerically, is required to properly understand the gas misalignment
in ETGs. Such work is also timely, since if we can understand the
typical timescale over which misalignments are visible, new large
integral-field spectroscopic surveys of galaxies (e.g.\ the Calar Alto Legacy Integral Field
spectroscopy Area survey (CALIFA); \citealt{2015A&A...576A.135G}, the Sydney-AAO Multi-object Integral field
spectrograph (SAMI) galaxy survey, \citealt{Bryant2015}; the Mapping
Nearby Galaxies at APO (MaNGA) survey, \citealt{Bundy2015}) will
have the power to investigate these effects as a function of galaxy
mass, environment, etc, and thus
to accurately constrain the (gas-rich) merger rate in the local
universe.

\section*{Acknowledgments}
We thank the members of the ATLAS$^{\rm 3D}$ team for generating the
data at the origin of this paper and for constructive discussions. We
also thank Sam Geen, Claudia Lagos, Sugata Kaviraj and Freeke van de
Voort for useful exchanges. TAD acknowledges support from a Science
and Technology Facilities Council Ernest Rutherford Fellowship. MB
acknowledges support from STFC rolling grant `Astrophysics at Oxford'
PP/E001114/1 and the hospitality of Nagoya University while much of
this work was carried out. The research leading to these results has
received funding from the European Community's Seventh Framework
Programme (/FP7/2007-2013/) under grant agreement No 229517.
%
%
\bibliography{bibTimescales}
\bibliographystyle{mn2e}
\end{document}